# From product recommendation to cyber-attack prediction: Generating attack graphs and predicting future attacks

Nikolaos Polatidis[1] - Elias Pimenidis[2] - Michalis Pavlidis[1] - Spyridon Papastergiou[3] - Haralambos Mouratidis[1]

[1]School of Computing, Engineering and Mathematics, University of Brighton, BN2 4GJ, Brighton, United Kingdom
Nikolaos Polatidis: N.Polatidis@Brighton.ac.uk
Michalis Pavlidis: M.Pavlidis@Brighton.ac.uk
Haralambos Mouratidis H.Mouratidis@Brighton.ac.uk

[2]Department of Computer Science and Creative Technologies, University of the West of England, BS16 1QY, Bristol, United Kingdom
Elias Pimenidis: Elias.Pimenidis@Uwe.ac.uk

[3]Department of Informatics, University of Piraeus, 18534, Piraeus, Greece
Spyridon Papastergiou: Paps@Unipi.gr

**Abstract.** Modern information society depends on reliable functionality of information systems infrastructure, while at the same time the number of cyber-attacks has been increasing over the years and damages have been caused. Furthermore, graphs can be used to show paths than can be exploited by attackers to intrude into systems and gain unauthorized access through vulnerability exploitation. This paper presents a method that builds attack graphs using data supplied from the maritime supply chain infrastructure. The method delivers all possible paths that can be exploited to gain access. Then, a recommendation system is utilized to make predictions about future attack steps within the network. We show that recommender systems can be used in cyber defense by predicting attacks. The goal of this paper is to identify attack paths and show how a recommendation method can be used to classify future cyber-attacks in terms of risk management. The proposed method has been experimentally evaluated and validated, with the results showing that it is both practical and effective.

## 1. Introduction

Recommender systems are decision support systems available on the web to assist users in the selection of item or service selection in online domains. In doing so recommender systems assist users in overcoming the information overload problem (Lu, Wu, Mao, Wang, & Zhang, 2015; Polatidis & Georgiadis, 2013). Collaborative filtering (CF) is the most widely used method for providing personalized recommendations. In CF systems, a database of user submitted ratings is used and the generated recommendations are generated on how much a user will like an unrated item based on previous common rated items. Thus, the recommendation process is based on assumptions about previous rating agreements and if these agreements will be maintained in the future. In addition, the ratings are used to create an n x m matrix with user ids, item ids and ratings, with an example of such a matrix shown in table 1. This database has four users and four items with values from 1 to 5. The matrix is used as input when a user is requesting recommendations and for a recommendation to be generated the degree of similarity between the user who makes the request and the other users' needs to be predicted using a similarity function such as the Pearson Correlation Similarity (PCC) (Su & Khoshgoftaar, 2009). At the next step, a user neighborhood which consists of users having the highest degree of similarity is created with the requester. Finally, a prediction is generated after computing the average values of the nearest neighborhood ratings about an item, resulting in a recommendation list of items with the highest predicted rating values.

Table 1. An Example of a Ratings Matrix

|        | Item 1 | Item 2 | Item 3 | Item 4 |
|--------|--------|--------|--------|--------|
| User 1 | 1      | 2      | 5      | -      |
| User 2 | 4      | 5      | 4      | 1      |
| User 3 | -      | -      | 3      | 2      |
| User 4 | 1      | 1      | 2      | 5      |

Even though, recommender systems have been used for product or service recommendation, in the current era where cyber-attacks have been increasing we show that they can be useful in attack prediction as well. In networks is important to be able to identify potential attacks made by local or network-based attackers and prevent them. Moreover, cyber-attackers tend to exploit vulnerabilities within a network and form attack paths from one asset to another until they have reached the asset they wish to harm. Recommender systems is a technology that has been used mostly in e-Commerce for product recommendation but can also be used in cyber-security to predict how an attacker might move within a network after a vulnerability has been exploited. Furthermore, it is a fact that among assets there exist common or similar vulnerabilities and a recommender system can be used to identify such similarities.

### 1.1 Problem definition and contributions

Cyber-attack prevention methods are based on graph analysis to identify attack paths or use previous attacker knowledge in combination with intrusion alerts to provide defense actions in real time. A gap is identified in attack prediction which can be solved with the use of suitable technologies. We have made the following contributions:

1. We identify all attack paths in a graph according to constraints.
2. We use the attack paths in combination with common vulnerability data to predict future attacks.
3. We use real data and a risk management system for the maritime supply chain IT infrastructure for the evaluation where we show that the method is both practical and effective.

### 1.2 Paper structure

In section 2 relevant background work is analyzed. In section 3 the proposed method is explained. Section 4 presents the experimental evaluation of the attack path discovery method, Section 5 is the evaluation of the attack prediction method, Section 6 is the discussion and section 7 contains the conclusions and future work parts.

## 2. Background

### 2.1 Attack graph generation and analysis

Cyber-attack prevention technologies typically use attack graph generation and analysis methods to identify all possible paths that attackers can exploit to gain unauthorized access to a system (Ou & Singhal, 2011). There are numerous methods available for attack graph generation and analysis. In (Templeton & Levitt, 2000) the authors use a general graph model, which is based on the JIGSAW specification language. Sample attack scenarios are created using different methods such as substitution, distribution and looping. In (Ning & Xu, 2003) the authors developed an intrusion correlator for intrusion alerts, which produces correlation graphs as output. Then, they use these graphs to create attack strategy graphs. The authors in (Ritchey & Ammann, 2000) utilize modeling based approach that is used to perform an analysis of the security of the network. This is done using model checking tools and a model is presented that describes the vulnerability to attack of the network. In (Sheyner, Haines, Jha, Lippmann, & Wing, 2002) the authors developed a tool called NuSMV, a Network Symbolic Model checker. This is a model checking tool that implements an algorithm for automatic generation of attack graphs. A logic-based approach is proposed in (Xinming Ou, Wayne F. Boyer, 2006). In this approach, the authors use logic rules to compute the attack graph and use logic deduction to reach the final facts from the initial facts. Although, this approach suffers from performance issues as the state grows. In (Ammann, Wijesekera, & Kaushik, 2002) a Breadth-first search solution is used by the authors to build the attack graph. A layered solution is proposed where the bottom layer contains attacker privileges and the upper layer contains the privileges computer after each step of the algorithm. Once again, as the size of the graph grows there are performance issues. In (Ammann, Pamula, Ritchey, & Street, 2005) the authors propose an algorithm that only creates a graph containing the worst case scenarios. This approach performs better in terms of performance, but it cannot guarantee that all relevant paths will be returned. In (Ingols, Lippmann, & Piwowarski, 2006) the authors try to reduce complexity by introducing the concept of group reachability. This method uses a breadth first method and uses prerequisite graphs that express reachability conditions among network hosts. The authors in (Ingols, Chu, Lippmann, Webster, & Boyer, 2009) develop further the prerequisite graphs by adding information about client-side attacks, firewalls and intrusion detection. In (Kaynar & Sivrikaya, 2016) the authors use a distributed attack graph generation algorithm based on a multi-agent system, a virtual shared memory abstraction and hyper-graph partitioning to improve the overall performance of the system. The method is based on depth first search and it is shown that the performance is improved with the use of agents after a specific graph size. In (Xie, Zhang, Hu, & Chen, 2009) the authors use a bidirectional search method to generate the attack graph. They also apply a restriction about the depth of the search, which limits the algorithm from identifying less possible attacks. In (Ghosh & Ghosh, 2012) an approach that is based on artificial intelligence with the name Planner is applied to generate the attack graph. Customized algorithms are used to generate attack paths in polynomial time. In (Phillips & Swiler, 1998) the authors propose a graph-based approach to analyze vulnerabilities, that can analyze risk to a specific asset and examine possible consequence of an attack. In (Almohri, Watson, Yao, & Ou, 2016) the use of a probabilistic model is proposed. This model measures risk security, computes risk probability and considers dynamic network features. A somewhat different approach is proposed by the authors in (Bi, Han, & Wang, 2016). The use of dynamic generation algorithm is proposed, that returns the top K paths. Furthermore, it is not required to calculate the full attack graph to return the top attack paths. NetSPA is a network security planning architecture that very efficiently generates the worst case attack graphs (Artz, 2002). To do this the system uses information from software types and versions, intrusion detection systems, network connectivity and firewalls. In (Poolsappasit, Dewri, & Ray, 2012) the use Bayesian attack graph generation for dynamic security risk management. In (Ou, Govindavajhala, & Appel, 2005) the authors developed Multi-host, Multi-stage Vulnerability Analysis Language (MulVAL), a logic-based network security analyzer. This is a vulnerability analysis tool that models the interaction of software bugs along with network configurations. The data about the software bugs are provided by a bug-reporting community, while all the other relevant information is enclosed within the system. In addition to MulVAL, Topological Vulnerability analysis (TVA) is another tool for generating attack graphs (Jajodia, Noel, & O'Berry, 2005; Ou & Singhal,

2011). TVA is based on topological analysis of network attack vulnerability and the idea is to exploit dependency graph to represent preconditions and postconditions and then exploit. At the next step, a search algorithm finds attack paths that exploit multiple vulnerabilities.

**2.2 Collaborative filtering**

As explained above a database of ratings and a similarity function such as PCC are the two essential parts of the CF recommendation process. Except for the classical recommendation method, PCC, another similar method found in the literature is weighted PCC (WPCC) which extends PCC by setting a statically defined threshold of common rated items. However, since the definitions of PCC and WPCC numerous approaches have been proposed with the aim of improving the recommendations. TasteMiner is a method that efficiently mines rating for learning partial users tastes to restrict the neighborhood size, thus reducing complexity and improving the accuracy of the recommendations (Shams & Haratizadeh, 2017). Another CF approach that aims to improve the accuracy of the recommendations is entropy based can be found in the literature. In this approach an entropy driven similarity used to calculate the difference between ratings and a Manhattan distance model is then used to address the fat tail problem (Wang, Zhang, & Lu, 2015). One more similarity measure for improving the accuracy of CF has been proposed with the name PIP. This measurement is based on Proximity, Impact and Popularity (PIP). Initially the proximity factor is applied to calculate the absolute difference between two ratings, then the impact factor is applied to show how strongly an item is preferred and finally the popularity factor is applied to how common the user ratings are. These three factors are then combined to calculate a final value (Liu, Hu, Mian, Tian, & Zhu, 2014). HU-FCF is a hybrid fuzzy CF method for improved recommendations (Son, 2014). In this method, CF is extended with a fuzzy similarity that is calculated on user demographic data. A CF recommendation method based on singularities has been proposed (Bobadilla, Ortega, & Hernando, 2012). In this method, the traditional similarities can be improved if contextual information from the entire user body are used to calculate singularities. Thus, the larger the singularity between users then the impact of it in the similarity is larger. Additionally, the use or power law augments to similarity values can be found in the literature with the name PLUS (M. Gan & Jiang, 2013). PLUS, is a method applied to user similarities to adjust their value using a power function and achieves a tradeoff between accuracy and diversity of the recommendations. Yet another approach for improved recommendations is the use of Pareto dominance (Ortega, Sánchez, Bobadilla, & Gutiérrez, 2013). Pareto dominance is used initially as a pre-filtering service were the less promising users are eliminated from the user neighborhood. Then, the rest are used in a typical CF recommendation process. An additional recommendation approach includes the breakup of the user neighborhood in multiples levels (Polatidis & Georgiadis, 2016). This can be done either using a static approach or a dynamic one (Polatidis & Georgiadis, 2016, 2017). In both approaches the user similarities are adjusted either in a positive or a negative way based on the number of co-rated items and the PCC values and are assigned to one of multiple levels based on the final computed value. Thus, the predictions are made using the new user neighborhood and the recommendations are improved. An additional method that can be used to improve the quality of the recommendations is natural noise removal (Toledo, Mota, & Martínez, 2015). Items and users are characterized based on their profiles and a defined strategy is used to eliminate natural noise, thus receiving more accurate recommendations. Also, other traditional approaches exist that can be used to improve CF and include the use of content-boosted CF or the utilization of sparsity measures (Anand & Bharadwaj, 2011; Melville, Mooney, & Nagarajan, 2002). COUSIN is a recommendation model that improves both the accuracy and the diversity of the recommendations by using a regression model that effectively removes weak user relationships (M. Gan, 2016). There is also an approach in the literature called Trinity that uses historical data and tags to provide personalized recommendations based on a three-layered object-user tag network (M.-X. Gan, Sun, & Jiang, 2016). In addition to the methods mentioned already the use of user-item subgroups has been proposed as a way of providing improved recommendation systems (Xu, Bu, Chen, & Cai, 2012).

**2.3 Combination of attack graph analysis and collaborative filtering for attack prediction**
In sections 2.1 and 2.2 related works regarding attack graph analysis and collaborative filtering methods have been explained. In section 2.1 related works about graph analysis have been analysed due to the fact that analysing a graph and identifying possible attack paths is the first path of the attack path recommendation process. In section 2.2 related works about collaborative filtering recommendation methods have been analysed to identify the most relevant ones that can be used for attack prediction. The related works have been analysed in order to identify the most relevant attack graph analysis method and the most relevant recommendation method. We have most suitable methods from both categories that have the most features. For attack graph analysis a method has been selected that satisfies criteria such as the location and the knowledge of the attacker and supports pruning of paths. On the other hand collaborative filtering is the most suitable method since it provides a reliable method for identifying similar vulnerabilities. Furthermore, multi-level collaborative filtering works better when the similar vulnerabilities have very similar characteristics, since it considers the common similar vulnerabilities except the similarity values derived from similar vulnerabilities.

**3. Proposed method**
Our proposed method takes elements from both collaborative filtering recommender systems and attack path discovery methods to identify attacks paths and predict attacks. Initially, we use an attack path discovery method that has unique characteristics, such as the attacker location, the attacker capability and which the entry and target points are (Polatidis, Pavlidis, & Mouratidis, 2018; Polatidis, Pimenidis, Pavlidis, & Mouratidis, 2017). The attack path discovery method returns all non-circular attack paths that exist between assets that belong to the specified characteristics.

**3.1 Attack path discovery**
Attackers can use a set of basic privileges that can satisfy some initial input requirements to gain unauthorized access to a system. Attack graphs show every possible path that an attacker can use to gain further privileges (Barik & Mazumdar, 2014; Ou & Singhal, 2011). In general, various vulnerabilities, such as software vulnerabilities or inappropriate configuration settings, exist in information systems and can be exploited by attackers to gain access. An infrastructure it typically comprised of numerous nodes that can be exploited to intrude into the network. In addition, the number of vulnerabilities that exist on the network and the reachability conditions that occur are the factors that determine the size of the attack graph. In, addition as the graph becomes larger, the possibility of more exploitation options for an attacker increases. To build the attack graph we use direct conditions and utilize information from open sources. Initially, the weaknesses defined in the Common Weakness Enumeration (CWE) ("CWE," n.d.) are used, and at the second step, Information from the Common Vulnerabilities and Exposures (CVE) ("CVE," n.d.) database are used. A model is introduced where an attacker can gain access to information system sources and move in a directed path. Moreover, a set of preconditions are specified, which include the length of the path, the location and capability of the attacker.

The pseudocode of the attack path discovery is shown in algorithm 1, while the following activities need to be executed for the algorithm to identify the attack paths, while the term business partners refer to partners of a supply chain in the maritime sector:

1. **Activity 1: Entry Points Identification**: The Business Partners have to define the Entry Points (assets from which the attacks will be initiated; these assets are considered as more reachable by an attacker). Moreover, the business partners should be experts with an information technology (IT) and security background and be know the infrastructure.
2. **Activity 2: Target Points Identification**: The Business Partners must define the Target Points (the assets which are considered as target for attacks due to their criticality).
3. **Activity 3: Identify Attacker Profile**: Attacker profiles will be identified by their location and their expertise. Their location is represented by the values 1, 2 and 3 (local, adjacent and network). Their expertise is represented by the values 1, 2 and 3 (low, medium and high). In algorithm 1 the variables 'attacker location' and 'attacker capability' takes values from 1 to 3.
4. **Activity 4: Generate Vulnerability Chains**: This step follows a rule-based reasoning approach (filters) to generate the chain of sequential vulnerabilities on different assets that arise from consequential multi-steps attacks initiated from the Entry Points to exploit the vulnerabilities of the Target Points.

---

**Algorithm 1: Attack path discovery**

**Input:** Asset graph (G), attacker location, attacker capability
**Output:** Graph, affected assets, attack paths

*#We create two empty lists to hold attack paths and assets*
attackpaths = [] affectedassets = []
*#We return all paths from source to target*
**for** e in parameters entry points
    **If** (attacker location < required level of attacker location
    /*explain attacker location
    OR attacker capability < required attacker capability)
    /*explain attacker capability
    **return** empty graph
        **else if**
        (attacker location >= required level of attacker location
        OR attacker capability >= required attacker capability)
        AND
        (vulnerability type == Code execution
        OR vulnerability type == Code overflow
        OR vulnerability type == XSS
        OR vulnerability type == Bypass something
        OR vulnerability type == Obtain privilege
        OR vulnerability type == Memory corruption)
**get** single source shortest path length
**set** propagation length for entry point e
**for** target point t
*#Create a list with all non-circular paths from entry e to target t*
**get** all paths in the graph G from entry e to target t that are up to the pre-specified path length
    **for** the size of paths found
        **add** paths to attackpaths [] list, **add** affected assets to affectedassets [] list
*#Return the graph, the affected assets and the attack paths found as a direct input to*
*#the attack visualization algorithm*
**return** Graph, affected assets, attack paths

## 3.2 Attack prediction

To recommend attack predictions we use a parameterized version of multi-level collaborative filtering method described in (Polatidis & Georgiadis, 2016), although other methods could be applied according the scenario and the available data. This method applies collaborative filtering and then rearranges the order of the k nearest neighbors according to the similarity value and the number of co-rated items. We use characteristics from the above-mentioned method to classify attacks. To do that we initially apply classical collaborative filtering using PCC defined in equation 1. In PCC *Sim (a, b)* is the similarity of users *a* and *b*, $r_{a,p}$ is the rating of user a for product *p*, $r_{b,p}$ is the rating of user *b* for product *p* and *r´a, r´b* represent user's average ratings. *P* is the set of all products. At the next step, we check the similarity values returned by equation 1 and the number of co-rated vulnerabilities. Depending on the similarity value returned and the common vulnerabilities, we classify these attacks from very high to very low. Finally, we check if there are any attack paths between the assets before the classification process is finished. A detailed explanation of the steps can be found in algorithm 2 which provides the pseudocode of the attack prediction recommender system. Finally, it should be mentioned that classical collaborative filtering without any other parameters could be used but this could raise issues when many common vulnerabilities exist between assets due to the fact of a high returned similarity value.

$$PCC\ a,b = \frac{\sum p \in P(r_{a,p} - r´a)(r_{b,p} - r´b)}{\sqrt{\sum p \in P(r_{a,p} - r´a)^2} \sqrt{\sum p \in P(r_{b,p} - r´b)^2}} \quad (1)$$

---

**Algorithm 2: Attack prediction**

**Input:** attack paths, affected assets, vulnerabilities
**Output:** predicted attacks

---

#Vulnerabilities refers to common vulnerabilities between assets
**load** vulnerabilities (CVEs)
**load** vulnerability types (CWEs)
**apply** equation 1 using vulnerabilities as input
**get** similarity values
#If there are common vulnerabilities, then typically these receive the same score
#between assets, thus, resulting in absolute similarities
#Then we rearrange the order of the similarity by adding the number of co-rated items as a constraint
#classification refers to predicted attack classification, which is from very high to very low
**then** #n represents the number of co-rated items and x1, x2, x3 and x4 are fixed integers
    **if** n>=x1 && vulnerability belongs to the same type **then** classification == very high
        **else if** (n<x1 && n>=x2) vulnerability belongs to the same type && **then**
        classification == high
        **else if** n<x2 && n>=x3 **then** classification == Medium
        **else if** n<x3 && n>=x4 **then** classification == Low
    **else** classification == very low
**then**
    **get** attack paths
        **if** attack path exists
            **set** classification == very high
**else if** attack path does not exist && classification == very high then classification == high
        **else** classification == classification
**Return** predicted attacks

## 4. Attack path generation evaluation

The experiments took place in a simulated environment using a Pentium i7 2.8 GHz with 12 gigabytes of RAM, running windows 10. Section 4.1 represents the performance evaluation, while section 4.2 provides a comparison with other methods.

## 4.1 Performance evaluation

This section presents the performance evaluation of the attack graph generation algorithm, with the results presented in table 2 and figure 1. To examine the performance and feasibility of the proposed attack paths generation approach to identify and calculate all the possible attack patterns, we will use the Port's Services Requested Supply Chain sub process of the "vehicles transport service", which is a part of the supply chain that includes a number of assets regarding this service only. The "Port's Services Requested" sub-process aims to illustrate the interactions among the Port Authority, the Ship Agent and the Customs to request services for the vessel's arrival or departure. Manifest Registration Number (MRN) is required in the current sub-process to precede with these tasks and includes assets regarding this service of the supply chain only. The Ship Agent submits the Manifest Registration Number (MRN) received from the Customs to the Port Authority requesting services for the vessel such as, mooring, lacing, assigning risk assessment processes weather conditions, navigational warnings, procedures for communication failure, fenders, personnel (truckers for transferring the vehicles from the Industry to storage area, etc.). This process is performed via the Port Community System (PCS) that is an electronic platform which connects the multiple systems operated by a variety of organizations involved in the port's supply chain. This system facilitates the secure and efficient electronic exchange of information between the public and private stakeholders and allows the automatization and the smooth operation of the port and logistics processes through a single request submission. It should be noted that about 180 cyber assets (35 hardware assets and 145 software assets) with different product characteristics and technical specifications (such as product version, vendor) as well as with several associated confirmed Vulnerabilities and flaws (including CVE's, CVSS data, vulnerability type or vulnerability details) identified that are necessary to support the provision of the process.

**Table 2.** Performance evaluation results

| No. of test | Attacker Capability | Propagation length | No of entry points | No of target points | Running time |
|---|---|---|---|---|---|
| 1 | Low | 3 | 5 | 5 | <1 |
| 2 | Low | 4 | 5 | 5 | <1 |
| 3 | Low | 5 | 5 | 5 | <1 |
| 4 | Medium | 3 | 5 | 5 | <1 |
| 5 | Medium | 4 | 5 | 5 | <1 |
| 6 | Medium | 5 | 5 | 5 | 1 |
| 7 | High | 3 | 5 | 5 | <1 |
| 8 | High | 4 | 5 | 5 | 1 |
| 9 | High | 5 | 5 | 5 | 1.2 |
| 10 | High | 3 | 25 | 25 | 1.35 |
| 11 | High | 5 | 25 | 25 | 1.50 |
| 12 | High | 10 | 25 | 25 | 1.95 |

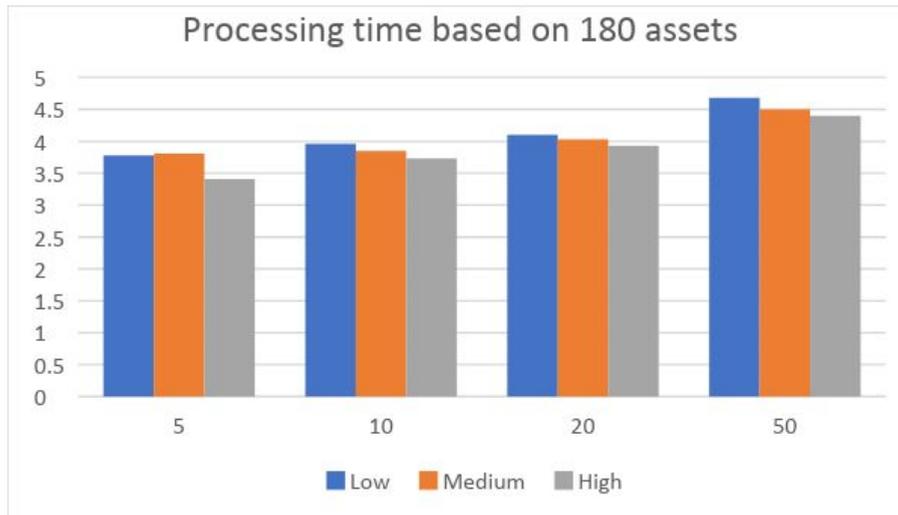

**Fig. 1.** Performance evaluation based on 180 assets

**4.2 Attack graph generation algorithm comparison with other methods**
The proposed method has been compared with the following long established and state-of-the-art alternatives. The first two are the most relevant and well-established methods found in the literature, while the following two are the most relevant state-of-the-art similar methods found in the literature. The methods have been selected due to the fact that are the most relevant and can be used for risk management. Moreover, the selected methods include both traditional and state-of-the-art approaches that satisfy most of the criteria discussed in section 4.2.1 and appendix A, while none of these alternatives satisfies the criteria as a whole.

1. Long established methods
   (Sheyner, Haines, Jha, Lippmann, & Wing, 2002 ).
   This is a model checking tool that implements an algorithm for automatic generation of attack graphs with the name NuSMV.
   (Jajodia et al., 2005; Ou & Singhal, 2011).
   This is a tool named TVA that is based on topological analysis of network attack vulnerabilities and the idea is to exploit a dependency graph to represent preconditions and postconditions and then exploit them. At the next step, a search algorithm finds attack paths that exploit multiple vulnerabilities.

2. State-of-the-art
   (Kaynar & Sivrikaya, 2016).
   This method uses a distributed attack graph generation algorithm based on a multi-agent system, a virtual shared memory abstraction and hyper-graph partitioning to improve the overall performance of the system.
   (Bi, Han, & Wang, 2016 ).
   The use of dynamic generation algorithm is proposed in this method and returns the top K paths.

**4.2.1 Evaluation criteria**
17 criteria have been identified and used for evaluating the quality of the algorithm. The selection was based on algorithm characteristics found on previous studies and current trends in risk management

(Kaynar & Sivrikaya, 2016; Lever & Kifayat, 2016; Polatidis et al., 2018; Yi et al., 2013). The details, of the criteria can be found in appendix A, while the criteria are also presented briefly within table 3.

### 4.2.2 Evaluation results
The results of the comparison are presented in table 3, where it is shown which of the criteria are satisfied by each method.

Table 3. Comparison results

| Criteria | Attack graph generation methods | | | | |
|---|---|---|---|---|---|
| | (Sheyner, Haines, Jha, Lippmann, & Wing, 2002) | (Jajodia et al., 2005; Ou & Singhal, 2011) | (Kaynar & Sivrikaya, 2016) | (Bi, Han, & Wang, 2016 | Proposed method |
| 1 (Attack Path Analysis) | √ | √ | √ | √ | √ |
| 2 (Vulnerability Chain Analysis) | × | √ | √ | × | √ |
| 3 (Integration of Open Source Information) | × | √ | √ | √ | √ |
| 4 (Integration of Crowd Sourcing Information) | × | × | × | × | √ |
| 5 (Collaboration Capabilities) | × | × | √ | × | √ |
| 6 (Support tool) | √ | √ | × | × | √ |
| 7 (Tool availability) | √ | × | × | × | √ |
| 8 (Pruning of paths) | × | × | √ | √ | √ |
| 9 (Propagation length) | × | × | × | √ | √ |
| 10 (Attacker location) | × | √ | √ | √ | √ |
| 11 | × | √ | √ | √ | √ |
| (Attacker capability) | | | | | |
| 12 (Entry points) | × | × | × | × | √ |
| 13 (Target Points) | × | × | × | × | √ |
| 14 (Satisfaction of EU policies) | × | × | × | × | √ |
| 15 (Can be used for risk assessment) | × | × | √ | × | √ |
| 16 (Vulnerability types) | × | √ | √ | √ | √ |
| 17 (Clarity and replication) | × | √ | √ | √ | √ |

The main goal of the attack path discovery method is to identify the attack paths in specified network fragments of the maritime supply chain infrastructure and use them for risk management. Furthermore, the attack path discovery method includes the satisfaction of the following important components, that the related works fail to address as a whole:

1. Capability and location of the attacker.

2. Propagation length.

3. Entry and target points.

4. Pruning of paths

**5.** Satisfaction of EU policies

## 5 Attack prediction algorithm evaluation

The maritime supply chain infrastructure it typically comprised of numerous assets that can be exploited to gain access and reach specific assets by popping from one to another. For the case study, we have used a snippet of data derived from the Valencia port IT infrastructure. In table 4 the data used show the common vulnerabilities between assets and their respective score. Assets 1, 2 and 3 are hardware assets, while the description column represents the vulnerable software asset that is installed on the respective hardware asset. Furthermore, the assets and attacks paths between them are a vital part of risk assessment. The following non-circular attack paths are present in the system:

Asset1 → Asset2
Asset2 → Asset3
Asset2 → Asset1

However, it should be noted that attack paths might vary according to the specific settings used, such as the propagation length, attacker location, capability, entry and target points.

Table 4. Common vulnerabilities

| Assets | Description | CVE 2015-1769 | CVE 2015-2423 | CVE 2015-2433 | CVE 2015-2485 |
|---|---|---|---|---|---|
| Asset 1 (Desktop PC) | Windows 10 Installed on Desktop PC | 10 | 2.9 | 2.9 | 10 |
| Asset 2 (Laptop 1) | Windows 10 Installed on Laptop 1 | 10 | 2.9 | 2.9 | 10 |
| Asset 3 (Laptop 2) | Windows 10 Installed on Laptop 2 | 10 | 2.9 | 2.9 | - |

Then the administrator executed algorithm 2 to predict very high and high classification attacks. Moreover, for the case study we have assigned the minimum number of co-rated items to be 3 for very high classification and 2 for high classification. Thus, algorithm 2 classified:

1. Asset1 → Asset2 as very high
2. Asset2 → Asset1 as very high
3. Asset1 → Asset3 as high
4. Asset3 → Asset1 as high
5. Asset2 → Asset3 as high
6. Asset3 → Asset2 as high

At the next step, the method checked for attack path relations between the assets and rearranged the classifications. Thus, the administrator received the following final predictions:

1. Asset1 → Asset2 as very high
2. Asset2 → Asset1 as very high
3. Asset2 → Asset3 as very high
4. Asset1 → Asset3 as high
5. Asset3 → Asset1 as high
6. Asset3 → Asset2 as high

### 5.1 Expert validation

For the validation of the attack prediction method, the opinions of five experts have been gathered for the validity of each of the six predictions and how these should be classified. The outcome of the attack prediction method has been validated separately by each of the experts, while they had access to the database with the assets and the CVE vulnerabilities. The experts were selected among people from the business partners and had to be experts in IT with internal knowledge of the system.

The following comments where received by the experts:

1. A path between assets with vulnerabilities that are the same or that belong to the same CWE category are more important and should be classified as very high. This validates the fact that paths 1 to 3 are classified as very high.
2. Agreed that paths 4 to 6 are of high importance at least.

All five experts that although it is important to predict moves within a network, is also important to know that a true expert will try to exploit every possible vulnerability and movie within after exploiting any other possible vulnerability, according to the type of access the attacker wants. Furthermore, the experts agreed that when an attacker exploits a certain vulnerability then at the next step they would try to exploit either the same vulnerability or a vulnerability of the same type. While, at the same time three out of the five experts mentioned that they would try to identify if they could exploit a vulnerability that would cause a higher damage to the system and then other that would cause lesser damage. Finally, all agreed it is vital to have a tool that can make importance predictions, then further evaluate with experts the predictions and provide mitigation solutions accordingly.

## 6 Discussion

Risk management is important for identifying risks in networks and propose mitigation solutions. Typically, risk management systems rely on the use of attack graph generation methods to identify attack paths and perform risk assessments. Although, in the literature there are several attack graph generation methods there aren't any that satisfy plenty criteria to make the process straightforward and produce results of higher quality. In the literature there are long established methods such as the ones in (Sheyner et al., 2002) and (Jajodia et al., 2005) that can be used for attack graph generation. However, these methods being old and have been carefully examined in an experimental setting show that do not support several characteristics required for risk assessment such as pruning of paths and the location and capability of a potential attacker. On the other hand there are state-of-the-art methods such as the one in (Kaynar & Sivrikaya, 2016) and (Bi, Han, & Wang, 2016 ) that satisfy many more of the criteria necessary for risk assessment but several modifications would be necessary for them to be used. Thus, a new method for attack graph generation in terms of risk management was necessary and has been developed. The proposed attack graph generation method performs well in terms of performance as shown in the outputs in table 2 and in figure 1. Moreover, it satisfied all seventeen identified criteria, which will make the process of risk assessment and mitigation straightforward. Furthermore, cyber-attack prediction systems are important in risk management to provide mitigation solutions. To do that the identification of possible attack scenarios and providing defensive solutions for assets protection are the two most important parts. Furthermore, it is important for this to take place within a reasonable amount of time. It is shown that within a small amount of time the attack path discovery method delivers the non-circular attack paths between assets. Furthermore, at the next stage a classification list is created that provides a prediction list of attack movement between assets. For example, the likelihood that an attacker who gained access to asset 1 to explore the possibility of gaining access to asset 4 is higher when compared to gaining access to either asset 2 or asset 3. However, the possibility of common vulnerabilities receiving different scores in different assets should be further exploited since this will result in different classification scales.

## 7. Conclusions and future work

Recommender systems have been used extensively in various on-line services for product or service recommendation. However, until this point the use of such systems for predicting cyber-attacks has not been explored. In this paper we provide an in depth analysis of why cyber-attack prediction is important and how attack graph analysis can be combined with a collaborative filtering based approach to predict attacks within a risk management system. The proposed method combines attack graph analysis and recommendation technologies. Initially, a network is analyzed and a graph containing relevant attack paths is produced and on the next step multi-level collaborative filtering is used to predict how an attacker could move after access is gained to any of the assets. Furthermore, both parts of the proposed method have been evaluated for performance and quality. While, the method is practical, it could become more effective if certain aspects are extended, thus in the future we aim to investigate the following research directions:

**Path length recommendation.** We aim to apply recommendation techniques to dynamically identify the length of the path that should be searched, thus making the attack path discovery process faster.

**Cyber-attack prediction.** We aim to further develop our recommendation based attack prediction algorithm using classification methods such as Naïve Bayes and random forests.

**Appendix A Evaluation criteria**

1. Attack Path Analysis.
➢ This describes the capacity of the evaluated method to identify and analyses different attack paths. We are distinguishing the following main types.

2. Vulnerability Chain Analysis.
➢ This describes the capacity of the methods to identify chains of sequential vulnerabilities on different assets and include them into the risk analysis. We are distinguishing the following main types.

3. Integration of Open Source Information.
➢ This describes the capacity of the evaluated method to retrieve and integrated information coming from openly accessible sources of information (e.g., open source databases).

4. Integration of Crowd Sourcing Information.
➢ This describes the capacity of the evaluated method to retrieve and integrated information coming from crowd sourcing (e.g., technical forums).

5. Collaboration Capabilities.
➢ This describes the capacity of the evaluated method to enable and utilize the collaboration of several users in the risk analysis or risk management process.

6. Supporting tool.
➢ If there is a tool for providing a visual representation or any other relevant form of the results.

7. Tool availability.
➢ If the tool is available to the public to download, use or modify.

8. Pruning of paths.
➢ Pruning of paths makes algorithm more efficient. The algorithm can cut paths that either not important or fall in a category that we are not interested in, such as networked attacks.

9. Propagation length.
➢ The propagation length can be specified. The user should be able to enter the length that a potential attacker could reach after gaining access to an entry asset.

10. Attacker location.
➢ The location of the attacker can be specified. The location of the attacker can be specified, and it should be either local or networked.

11. Attacker capability.
➢ The capability of the attacker can be specified. The capability should be specified in terms of high, medium, low or similar.

12. Entry points.
➢ The entry assets can be specified, which helps to search on specific network parts for problems.

13. Target points.
   The target assets can be specified, which helps to search on specific network parts for problems.

14. Satisfaction of EU policies.
➢ EU maritime supply chain policies are satisfied.

15. Can be used for risk assessment.
➢ This describes the applicability of the evaluated method for the maritime supply chain risk assessment area.

16. Vulnerability types.
➢ The types and the categories of the vulnerabilities can be specified within the settings of the algorithm.

17. Clarity and replication.
➢ The algorithm is presented in a manner that it makes it easy to replicate or extend.